\documentclass{aa}  
\usepackage{natbib}
\bibpunct{(}{)}{;}{a}{}{,} 
\usepackage{txfonts}
\usepackage{tasks}
\usepackage{float}
\usepackage{graphicx}
\usepackage{rotating}
\usepackage{booktabs}
\usepackage{dblfloatfix}
\usepackage{hyperref}
\usepackage[off]{auto-pst-pdf}
\usepackage{afterpage}
\usepackage{amsmath}
\graphicspath{ {images/} }

\begin{document} 

   \title{The milliarcsecond-scale radio structure of AB Dor A}


   \author{J.B. Climent
          \inst{1}\thanks{Email: j.bautista.climent@uv.es}
          \and
          J. C. Guirado\inst{1}\fnmsep\inst{2}
          \and
           R. Azulay\inst{1}\fnmsep\inst{3}
           \and
          J.M. Marcaide\inst{1}\fnmsep\inst{4}\thanks{Visiting Professor at the Department of Quantum Physics and Astrophysics, and the Institute of Cosmos Sciences of the University of Barcelona}
          \and
          D.L. Jauncey\inst{5}\fnmsep\inst{6}
           \and
          J.-F. Lestrade\inst{7}
                     \and
          J.E. Reynolds\inst{5}
          }

         \institute{Departament d’Astronomia i Astrof\'isica, Universitat de Val\`encia, C. Dr. Moliner 50, E-46100 Burjassot, Val\`encia, Spain
         \and
         Observatori Astron\`omic, Universitat de Val\`encia, Parc Cient\'ific, C. Catedr\'atico Jos\'e Beltr\'an 2, E-46980 Paterna, Val\`encia, Spain
         \and 
         Max-Planck-Institut f\"ur Radioastronomie, Auf dem H\"ugel 69, D-53121 Bonn, Germany
         \and 
         Donostia International Physics Center, Paseo de Manuel Lardizabal 4, E-20018 Donostia-San Sebasti\'an, Spain
         \and
         CSIRO Astronomy and Space Science, Australia Telescope National Facility, Epping, NSW 1710, Australia
         \and
          Research School of Astronomy and Astrophysics, Australian National University, Canberra, ACT 2611, Australia
          \and
          Observatoire de Paris, PSL Research University, CNRS, Sorbonne Universités, UPMC, 61 Av. de l’Observatoire, F-75014 Paris, France
         }

   \date{Received ...; accepted ...}

 
  \abstract
   {
   The fast rotator, pre-main sequence star AB Dor A is a strong and persistent radio emitter. The extraordinary coronal flaring activity is thought to be the origin of compact radio emission and other associated phenomena, such as large slingshot  prominences.
   }
   {
   We aim to investigate the radio emission mechanism and the milliarcsecond radio structure around AB Dor A.
   }
   {
   We performed phase-referenced VLBI observations at 22.3 GHz, 8.4 GHz, and 1.4 GHz over more than one decade using the Australian VLBI array. 
   }
   {
   Our 8.4 GHz images show  a  double  core-halo  morphology, similar at all epochs, with emission extending at heights between 5 and 18 stellar radii. Furthermore, the sequence of the 8.4 GHz maps shows a clear variation of the source  structure  within  the  observing  time.  However, images at 1.4 GHz and 22.3 GHz are compatible with a compact source. The phase-reference position at 8.4 GHz and 1.4 GHz are coincident with those expected from the well-known milliarcsecond-precise astrometry of this star, meanwhile the 22.3 GHz position is 4$\sigma$ off the prediction in the north-west direction. The origin of this offset is still unclear.} 
   {
   We have considered several models to explain the morphology and evolution of the inner radio structure detected in AB Dor A. These models include emission from the stellar polar caps, a flaring, magnetically-driven loop structure, and the presence of helmet streamers. We also investigated a possible close companion to AB Dor A. Our results confirm the extraordinary coronal magnetic activity of this star, capable
of producing compact radio structures at very large heights that have so far
only been seen in binary interacting systems.

   }


   \maketitle
%
\section{Introduction}\label{sect:introduction}

Young stellar objects (YSOs) exhibit radio emission as a result of a wide range of mechanisms. One of the most common mechanisms detected through observations \citep[see][and references therein]{2011ApJ...736...25F} occurs when electrons gyrate in the magnetic fields of these objects, producing non-thermal continuum emission called cyclotron, gyro-synchrotron, or synchrotron emission depending on the velocity of such electrons \citep{1985ARA&A..23..169D}.
The majority of YSOs with non-thermal radio emission that have been detected are pre-main-sequence (PMS) stellar objects with the notable exception of a few Class I protostars \citep{2006A&A...446..155F,2013A&A...552A..51D}. The radio emission in these objects is thought to originate in and remain confined to the magnetosphere \citep[with a typical size of a few stellar radii; see][]{2007prpl.conf..479B}. Consequently, it typically remains unresolved at the current very long baseline interferometry (VLBI) resolution (milli-arcsecond), as demonstrated by the largest existing sample of VLBI detections of PMS stars \citep{2017ApJ...834..141O}. 

In a few cases, VLBI observations of PMS stars have revealed resolved magnetospheres of sizes that are up to several times the stellar radii \citep{1991ApJ...382..261P,1991ApJ...376..630A,1992ApJ...401..667A}. Other objects exhibiting extended magnetospheres include RS CVn \citep{1988A&A...197..200M,2001A&A...373..181T,2002ApJ...572..487R,2003ApJ...587..390R}, Algol-like binaries \citep{1998ApJ...507..371M,2010Natur.463..207P}, M dwarfs \citep{1998A&A...331..596B}, and chemically peculiar Bp/Ap stars \citep{1988Natur.334..329P}. 
Based on some of these VLBI results, together with measured radio spectra and polarization measurements, a magnetospheric model consisting of a global dipole-like structure was suggested \citep[see Fig. 3 in][]{1988ApJ...335..940A}.  The radio emission detected in these objects is believed to come from synchrotron or gyrosynchrotron radiation from energetic electrons accelerated by magnetic reconnections. Most of the radio emission would originate in: i) the magnetic equator line at a distance where the stellar wind opens the closed magnetic field lines and creates a current sheet of plasma where electrons are continuously accelerated; ii) close to the polar caps where the wind flows almost unrestrained; and iii) a combination of the previous two places. A version of this model has successfully explained the two radio lobes of opposite circular polarization seen in Algol \citep{1998ApJ...507..371M} and UV~Cet~B \citep{1998A&A...331..596B}, where each lobe would be located in a different polar cap region.

Of particular interest to our work is the case of the young binary system V773 Tau A. This system exhibits a persistent radio emission that varies in intensity depending on the separation between components, indicating some interaction between the individual magnetospheres \citep{2012ApJ...747...18T}. More remarkably, the detected radio emission in this object may be explained by the presence of solar-like helmet streamers. Those streamers would be anchored to the top of a closed loop at a few stellar radii above the stellar surface. However, they would extend up to the upper mirror points located 30 stellar radii away \citep{2008A&A...480..489M}. Posterior observations did not confirm the detection of these mirror magnetic points \citep{2012ApJ...747...18T}; however, the interaction between individual magnetospheres was confirmed and the hypothesis of helmet streamers should be added to the list of possible scenarios occurring in these PMS stars.
                
Radio emission has also been detected in ultracool dwarfs (UCDs). The discovery of the first radio emission from an UCD proved the existence of powerful magnetic fields ($\sim$kG) on these objects \citep{2001Natur.410..338B}.
The presence of mildly relativistic electrons accelerating along the lines of these magnetic fields may produce what is known as electron cyclotron maser instability (ECMI) emission in a similar fashion to the auroral radio bursts observed in Solar System planets \citep{2001Ap&SS.277..293Z}. This emission is then detected as bright, highly-polarized radio bursts and, in some cases, with periodicity equal to the rotation period \citep{2009ApJ...695..310B,2010A&A...524A..15D,2013ApJ...779..101H,2014ApJ...788...23W}. 
Additionally, these electrons spiraling in an ambient magnetic field may produce non-bursting radio emission (quiescent emission) via the gyrosynchrotron mechanism. This emission varies on timescales of weeks (and longer) but it is typically steady over the observing time \citep{2007A&A...472..257A,2012ApJ...746...23M}. The only known exception is the most radio-bright UCD, NLTT 33370 B, which has demonstrated consistent level of radio emission for more than a decade \citep{2011ApJ...741...27M}.
These scenarios are not mutually exclusive and some of the known radio-active UCDs present both quiescent and bursting emission \citep{2005ApJ...627..960B,2006ApJ...653..690H,2008ApJ...684..644H,2015ApJ...799..192W}.
Understanding the phenomenology and the physical mechanisms involved in the radio emission of UCDs is crucial for planet discovery around them \citep{2014Sci...345..440R} and for investigating the atmosphere and habitability of these planets \citep{2015Sci...350.0210J,2016PhR...663....1S}, which could be abundant and observationally accessible \citep[e.g., TRAPPIST-1 system, see][]{2016Natur.533..221G,2017Natur.542..456G}. The number of UCDs found with detected radio emission has been increasing over the last years, with objects of spectral type as late as T6.5 \citep{2016ApJ...818...24K}, including a likely planetary-mass T2.5 object \citep{2018ApJS..237...25K}.

     About 15 pc away, AB Doradus (AB~Dor) is one of the most active and extensively studied PMS objects. It is a multiple system formed by two pairs of stars separated by 9$''$, AB~Dor~A/C and AB~Dor~Ba/Bb \citep{2005Natur.433..286C,2006A&A...446..733G}, which have lent their names to the AB Dor moving group (AB Dor-MG). The main star of this system, the K0 dwarf AB Dor A, 
        is a fast rotator (period of ~0.5 days; see Table \ref{table:properties}), which presents strong emission at all wavelengths, from radio to X-rays \citep[see][and references therein]{2019A&A...628A..79S}. It has been well studied by the \textit{HIPPARCOS} satellite \citep{1995A&A...304..182L,1995A&A...304..189L} and VLBI arrays \citep{1995A&A...304..182L,1997ApJ...490..835G}. This joint effort has revealed the presence of AB~Dor~C, a low-mass companion with 0.090 \(M_\odot\), orbiting AB~Dor~A at an average angular distance of 0.2$''$. The pair AB~Dor~A/C has also been observed by different near-infrared instruments at the VLT \citep{2005Natur.433..286C,2007ApJ...665..736C,2008A&A...482..939B}, allowing for independent photometry of AB~Dor~C which, along with the dynamical mass determination, has served as a benchmark for stellar evolutionary models. Recent evidence indicates that AB~Dor~C may be a binary system itself \citep{2019ApJ...886L...9C}, consisting of two brown dwarfs, AB Dor Ca/Cb, with 72 and 13 Jovian masses, respectively.
        The exact age of the system is a current subject of discussion: 40–50 Myr for AB~Dor~A and 25–120 Myr for AB~Dor~C in \citet{2017A&A...607A..10A}, 40-60 Myr in \citet{2004ApJ...613L..65Z} and \citet{2006ApJ...643.1160L}, 30-100 Myr in \citet{2005Natur.433..286C}, 40-100 Myr in \citet{2005AN....326.1033N}, 75-150 Myr in \citet{2006ApJ...638..887L}, 50-100 Myr in \citet{2007A&A...462..615J} and \citet{2008A&A...482..939B}, 40-50 Myr in \citet{2011A&A...533A.106G}, >110 Myr for the AB~Dor nucleus star in \citet{2013ApJ...766....6B}, and 130-200 Myr in \citet{2015MNRAS.454..593B}. 
        
        \citet{1984MNRAS.208..865S} were the first to detect radio emission from AB~Dor, finding a clear modulation in its flux and an absence of circular polarization \citep{1986PASAu...6..312S}. They interpreted this modulation as spot groups on the star that may have lifetimes on the order of two years or more. A more detailed study of the modulation in radio emission \citep{1992ApJ...388L..27L} found that the peaks in emission coincided with the two stellar longitudes at which starspots preferentially form and detected a large spot at one of these active longitudes. These starspots were also the interpretation adopted by other authors who suggested a model for them based on optical photometry and spectroscopy \citep[see Fig. 5 in][]{2009AN....330..358B}. Subsequent monitoring of the AB~Dor~A/C system with the Australia Telescope Compact Array (ATCA) has shown a baseline flux of 2 mJy and flare events up to 8 mJy, with a half-brightness duration of about 3 hours \citep{2014PASA...31...21S}, indicating that the modulation is still present even 23 years after the discovery of radio emission from this star. Finally, from VLBI multiepoch observations of AB~Dor~A, \citet{2017A&A...607A..10A} reported both revised orbital elements and dynamical masses for the AB~Dor~A/C system. 
        
    \begin{table}
    \caption{Physical properties and orbital parameters of AB~Dor~A/C} 
    \label{table:properties} 
    \centering 
    \begin{tabular}{l c c} 
    \hline\hline 
    Parameter & Value & Ref. \\ 
    \hline
    Abs. position AB~Dor~A$^{a}$\\
    \quad $\alpha_0$ (h m s) &  5 28 44.79483 & 1\\ 
    \quad $\delta_0$ ($^{\circ}$ $'$ $''$) & -65 26 55.91774 & 1\\
    Parallax (mas) & 66.4 $\pm$ 0.3 & 1\\
    Distance (pc) & 15.06 $\pm$ 0.10 & 1\\
    Orbital elements AB Dor A$^{b}$\\
    \quad P (years) & 11.78 $\pm$ 0.10 & 1\\
    \quad $a_A$ (mas)& 31 $\pm$ 1 & 1\\
    \quad $e$& 0.59 $\pm$ 0.05 & 1\\
    \quad $i$ $(^{\circ})$& 65 $\pm$ 1 & 1\\
    \quad $\omega_A$ $(^{\circ})$& 114 $\pm$ 5 & 1\\
    \quad $\Omega$ $(^{\circ})$& 132 $\pm$ 2 & 1\\
    \quad $T_0$ & 1991.9 $\pm$ 0.2 & 1\\
    Relative position AB Dor C$^{c}$\\
    \quad Separation ($''$) & 0.156 $\pm$ 0.10 & 2\\
    \quad P.A $(^{\circ})$ & 127 $\pm$ 1 & 2\\
    Stellar properties\\
    \quad Spectral Type A & K0V & 3\\
    \quad Mass A ($M_\odot$) & 0.89 $\pm$ 0.08 & 1\\
    \quad Radius A ($R_\odot$) & 0.96 $\pm$ 0.06 & 5\\
    \quad Rot. Period A (days) & 0.51479 $\pm$ 0.00001 & 4\\
    \quad Spectral Type C & M8 & 2\\
    \quad Mass C ($M_\odot$) &  0.090 $\pm$ 0.008 & 1\\
    \hline 
    \end{tabular}
    \begin{flushleft}
    \footnotesize{\textbf{Notes.}\\
    (a) The reference epoch for AB~Dor~A is 2000.0. \\
    (b) We consider the pair AB~Dor~A and AB~Dor~C with the origin of coordinates placed at the center of mass of this system. P, $a_A$, $e$, $i$, $\omega_A$, $\Omega$, and $T_0$ represent the period of the orbit, the semi-major axis of AB Dor A apparent orbit in the sky, the ellipticity of the orbit, the inclination of the orbit on the sky, the longitude of the periastron of the orbit, the longitude of the ascending node, and the time of periastron passage, respectively.\\
    (c) The values of AB Dor C orbital parameters are given for epoch 2004.093.\\}
    \rule{0pt}{3ex}    
    \footnotesize{\textbf{References.} (1) \citet{2017A&A...607A..10A};  (2) \citet{2005Natur.433..286C}; (3) \citet{2006A&A...460..695T}; (4) \citet{1988MNRAS.235.1411I}; (5) \citet{2011A&A...533A.106G}
    }
    \end{flushleft}
    \end{table}
        
        
        \begin{table*}[t]
    \caption{Journal of observations} 
    \label{table:obs} 
    \centering 
    \begin{tabular}{c c c c c c} 
    \hline\hline 
    Frequency (GHz) & Observing Date & Array Configuration$^{\mathrm{a}}$ & UT range & Beam Size (mas) & P.A ($^{\circ}$)\\ 
    \hline
    8.4 & 11 Nov. 2007 & At, Cd, Hh, Ho, Mp, Pa & 10:00-22:00 & 2.5 $\times$ 0.9 & $-$3\\
    8.4 & 25 Oct. 2010 & At, Cd, Ho, Mp, Pa & 11:00-23:00 & 3.2 $\times$ 2.8 & $-$26\\
    8.4 & 16 Aug. 2013 & At, Cd, Hh, Ho, Mp, Pa, Ti, Ww & 15:00-03:00 &  2.2 $\times$ 0.7 & 4\\
    22.3 & 14 Jun. 2017 & At, Cd, Ho, Mp, Pa & 20:40-07:40 & 4.7 $\times$ 3.4 & $-$74\\
    1.4 & 6 Feb. 2018 & At, Ho, Cd, Pa & 03:24-14:00 & 19.8 $\times$ 15.3 & $-$55\\
    \hline 
    \end{tabular}
    \begin{flushleft}
    \footnotesize{
    \textbf{Notes.} 
    $^{\mathrm{a}}$ Australia Telescope Compact Array (At), Hobart (Ho), Ceduna (Cd), Hartebeesthoe (Hh), Mopra (Mp), Parkes (Pa), DSS43 -- NASA’s Deep Space Network Tidbinbilla (Ti), and Warkworth (Ww).} 
    \end{flushleft}
    \end{table*}
        
        Here we present the study of new 1.4 GHz and 22.3 GHz VLBI observations of AB~Dor~A. In addition, we re-analysed the 8.4 GHz VLBI data from \citet{2017A&A...607A..10A} and found an extended, fast-evolving coronal structure around AB~Dor~A. Our field of view allowed us to also probe the UCD binary AB~Dor~Ca/Cb and place strong upper limits to the radio emission in this object.
        We describe the observations in Section 2 and the data reduction and analysis in Section 3. Then we present the results from this analysis in Section 4. In Section 5, we discuss our results, considering different scenarios that might explain them. Finally, in Section 6, we present our conclusions.

\section{Observations}\label{sect:observations}

    We observed the binary system AB~Dor~A/C using the Australian Long Baseline Array (LBA) at 1.4 GHz (21 cm) and at 22.3 GHz (1.3 cm).
    On February 6th, 2018, we observed at 1.4 GHz for a total duration of 10 hours. The system was observed in phase referencing mode using the source BL Lac PKS 0516-621 (about 3.6$^{\circ}$ away) as a phase calibrator. The sequence calibrator-target lasted 3.5 minutes (2.5 minutes on source and 1 minute on the calibrator). Both right and left circular polarizations were recorded using four 16 MHz bandwidth subbands per polarization. The same strategy was followed with the 22.3 GHz LBA observations carried out on June 14th, 2017 with the same phase-referencing calibrator but with  a sequence calibrator-target lasting 3 minutes (2 minutes on source and 1 minute on the calibrator).

    Previous observations of the AB~Dor system at 8.4 GHz, already reported in  \citet{2017A&A...607A..10A}, were reanalyzed along with the new 1.4 GHz and 22.3 GHz data. See Table \ref{table:obs} for further details of these observations. 

\section{Data reduction and imaging procedure}\label{sect:data_red_and_analysis}

\begin{figure*}[h]
    \centering
    \includegraphics[width=0.95\linewidth]{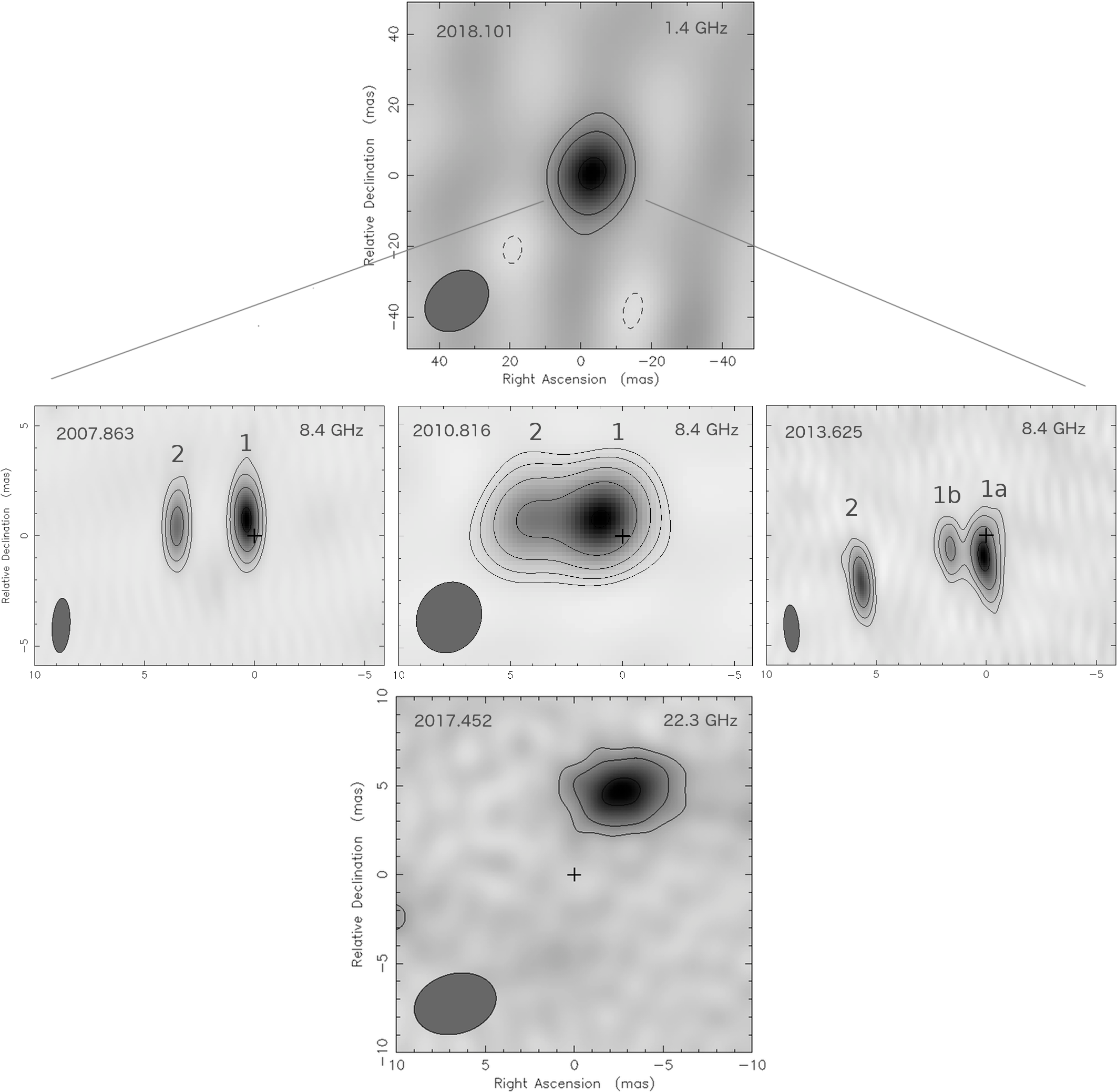}
    \caption{LBA images of all our observations of AB Dor A. Here and hereafter, north is up and east is to the left. Numbers (if any) indicate the index assigned to each component. For each image, the  
    synthesized beam is shown at the bottom left corner (see Table \ref{table:obs}). The lowest contour level (the remaining levels are spaced by factors of 2), the peak brightness, and background rms noise for each image are as follows: \textit{(2007)}: 11\%, 2.35  mJy beam$^{-1}$, 0.06 mJy beam$^{-1}$; \textit{(2010)}:  6\%, 2.16  mJy beam$^{-1}$, 0.05 mJy beam$^{-1}$; \textit{(2013)}:  10\%, 1.80  mJy beam$^{-1}$, 0.05 mJy beam$^{-1}$; \textit{(2017)}:  22\%, 0.714  mJy beam$^{-1}$, 0.05 mJy beam$^{-1}$; \textit{(2018)}:  22\%, 5.84  mJy beam$^{-1}$, 0.6 mJy beam$^{-1}$. All the images are centered, and cross-marked, at the expected positions of AB Dor A according to its proper motion, parallax, and orbital wobble.}
    \label{fig:allepochs}
    \end{figure*}

    We reduced the data shown in Table \ref{table:obs} using the Astronomical Image Processing System (AIPS) of the National Radio Astronomy Observatory (NRAO) following standard routines. Firstly, we calibrated the ionospheric delay and corrected the instrumental phases. We then calibrated the visibility amplitudes using the nominal sensitivity for each antenna and corrected the phases for parallactic angles. Finally, we performed a fringe-search on the phase calibrator to minimize the residual contributions to the phases and applied these new corrections to our target. 
    The phase-referenced, channel-averaged images were obtained using the Caltech imaging program DIFMAP \citep{1994BAAS...26..987S} with the clean algorithm while selecting the polarization of interest in each case. 
    We firstly centered a box at the maximum emission peak and used the DIFMAP task \textit{clean}. Afterwards, if any additional flux was still present in the image above the noise level then we repeated the procedure by centering an additional box at the new maximum peak. We repeated these steps until no new peaks were distinguishable from the noise. This procedure was done interactively.
    The AB Dor A image for each epoch and frequency is shown at Fig. \ref{fig:allepochs}.
    
    In addition to producing an image of AB~Dor~A/C for each LBA dataset, we also fitted circular Gaussians to the interferometric visibilities (\textit{$uv$} plane) using the DIFMAP task \textit{modelfit}. 
    To estimate the errors in the fitted parameters (i.e., deconvolved size, density flux of each component, and distance between components), we followed the expressions described in \citet{1999ASPC..180..301F}. The fitting results are shown in Table \ref{table:modelfit}.
    
    With regard how our images compare with previous VLBI observations of AB~Dor~A, we reference  
    \citet{2017A&A...607A..10A}, who imaged for the first time AB~Dor~A at VLBI resolution. 
    Their main objective was astrometric, providing precise positions for the peak emission of AB~Dor~A.
    In our analysis, we used a different weighting scheme to gain more sensitivity (at the expense of a worse resolution) and no self-calibration. Following this procedure, we were able to resolve the internal structure of AB Dor A, finding several substructures (see Fig.~\ref{fig:allepochs}). 
    We note that we are able to reproduce the images presented in \citet{2017A&A...607A..10A} by following the procedure described therein.
    Therefore, the images presented in this work supersede those reported in \citet{2017A&A...607A..10A}.

    
    \begin{table}
    \centering
    \caption{Model components of the fit of circular Gaussians on the uv-plane for VLBI observations.}
    \label{table:modelfit}
    \begin{tabular}{ccccc} 
    \hline\hline
    Epoch & Comp.$^{\mathrm{a}}$ & $S$ & $\theta$ & T$_{\mathrm{b}}$\\
    & & (mJy) & (mas) & ($10^{5}$K)\\
    \hline
    2007.863 & 1 & 3.4 $\pm$ 0.5  & 1.41 $\pm$ 0.18 & 420\\
    & 2 & 1.3 $\pm$ 0.2 & 0.60 $\pm$ 0.10 & 890\\
    \hline
    2010.816 & 1 & 2.31 $\pm$ 0.17 & 1.30 $\pm$ 0.09 & 340\\
    & 2 & 1.20 $\pm$ 0.16 & 1.51 $\pm$ 0.18 & 130\\
    \hline
    2013.625 & 1a & 2.7 $\pm$ 0.2 & 0.90 $\pm$ 0.08 & 830\\
    & 1b & 0.80 $\pm$ 0.15 & 0.40 $\pm$ 0.07 & 1200\\
    & 2 & 1.48 $\pm$ 0.19 & 0.65 $\pm$ 0.08 & 870\\
    \hline
    2017.452$^{\mathrm{b}}$& 1 & 1.2 $\pm$ 0.6 & 4.9 $\pm$ 1.3 & 1.5\\
    \hline
    2018.101 & 1 & 4.8 $\pm$ 0.8 & 11 $\pm$ 2 & 350\\
    \hline
    \end{tabular}
    \begin{flushleft}
    \footnotesize{\textbf{Notes.} $^{\mathrm{a}}$ We adopt the convention that the central component will be denoted by subindex 1. In case of detection, the subindex 2 will indicate the presence of a second component to the east. In 2013, subindices 1a and 1b indicate the central and the closest component to the east, respectively, while subindex 2 the furthest one. \textit{S} represents the flux density, \textit{$\theta$} the deconvolved FWHM diameter of the circular Gaussian component and T$_\mathrm{b}$ the minimum brightness temperature.
    $^{\mathrm{b}}$ 22.3 GHz data show a low S/N ($\sim$5) and should be treated carefully.
    }
    \end{flushleft}
    \end{table}
    
\section{Results}\label{sect:results}

\subsection{VLBI imaging and model fitting of AB Dor A}\label{sect:imaging_and_modelfitting}

    Fig.~\ref{fig:allepochs} shows the maps of AB~Dor~A at all observed frequencies and epochs. At 8.4 GHz, the brightest peak of emission is located at the map center, coincident (to within one beam size) with the expected position of AB~Dor~A, according to the kinematics reported in  \citet{2017A&A...607A..10A}.
    In Fig.~\ref{fig:allepochs}, the 8.4 GHz images show a complex structure around AB~Dor~A \citep[which was unnoticed in][see Section 5]{2017A&A...607A..10A}. 
    In 2007 and 2010, two emission peaks or components can be identified, separated by 3.1 $\pm$ 0.2 mas ($\sim$10 R$_\mathrm{star}$) and clearly oriented east-west. In the latter epoch, the double structure is not as clearly separated as in 2007 resulting in a double core-halo morphology. 

Later on, in 2013, the double structure can still be seen (separation of 5.7 $\pm$ 0.3 mas, $\sim$18 $R_\mathrm{star}$) although component 1 seems to be split in two close features, labeled as 1a and 1b in Fig. \ref{fig:allepochs}. However, we should remark that our visibility amplitudes and phases are similarly well-fitted by a component 1a, with a slight elongation toward the east. The reality of 1b as an independent feature should be taken with caution.
    The possible nature of the detected structures at 8.4 GHz is discussed in Sect. \ref{sect:discussion}.
    


    At 1.4 GHz, AB~Dor~A appears as a clear compact source with an estimated size of 11 $\pm$ 2 mas, whose position is also coincident with what is expected by the orbit determination of \citet{2017A&A...607A..10A}. Our 1.4 GHz data show no circular polarization (with an upper limit of 10\%) and minimum brightness temperature $\gtrsim$ 10$^{7}$ K, 
indicating non-thermal radio emission. 

    Finally, at 22.3 GHz (Fig. \ref{fig:allepochs}), AB~Dor~A shows an extended component located 5.5 mas away (17.5 $R_\mathrm{star}$) from the expected orbital position (marked with a cross). No significant circular polarization is found and the minimum brightness temperature is $\sim$10$^{5}$ K. Considering the low signal-to-noise ratio (S/N $\sim$5) and broad disagreement between the expected and measured peak positions, we consider that the validity of this detection may require further confirmation. 
    No background sources are expected at the peak position. The origin of this emission, if real, remains unclear.

\subsection{Time analysis of AB Dor A images}\label{sect:time_images}

        \begin{figure*}[h!]
    \centering
    \includegraphics[width=0.95\textwidth]{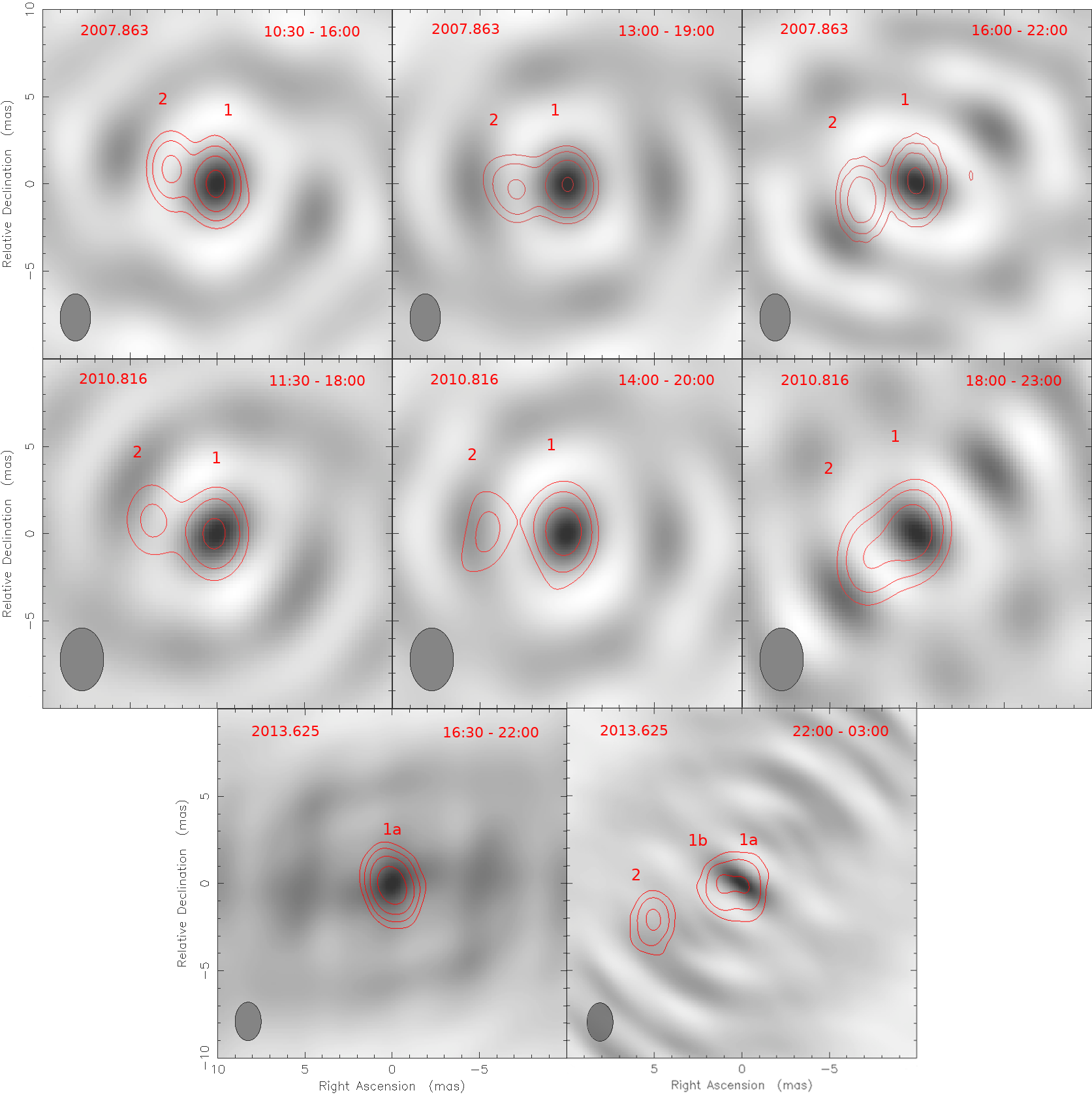}
    \caption{LBA snapshot images of AB Dor A at 8.4 GHz (as defined in the text) in red contours, while the dirty beam of the corresponding snapshot is shown in grayscale.
    Different epochs are shown in different rows. 
    Within the same epoch, the contour levels are the same and the beam size represents the average of the beam sizes of all the snapshots of such epoch, with 0$^{\circ}$ inclination for easier visualization. 
    The full width at half maximum (FWHM) beam size, its orientation and lowest contour level (the remaining levels are spaced by factors of 2) for each epoch are as follows: \textit{(2007)}: 1.7 $\times$ 2.7 mas at 0$^{\circ}$, 10\% of 3.01  mJy beam$^{-1}$; \textit{(2010)}: 2.5 $\times$ 3.6 mas at 0$^{\circ}$, 15.5\% of 2.46  mJy beam$^{-1}$; \textit{(2013)}: 1.5 $\times$ 2.2 mas at 0$^{\circ}$, 8\% of 3.84  mJy beam$^{-1}$.
    }
    \label{fig:image_time}
    \end{figure*}
    
    The images shown in Fig. \ref{fig:allepochs} correspond to the structure of AB~Dor~A 
    obtained with the full interferometric data set extending throughout the 
    complete duration of each observation (typically 10\,hr), which actually covers nearly 
    one rotation period of the star ($\sim$12\,hr).
If the morphology of AB~Dor~A happens to vary within the duration time of each observation, 
the observed structure seen at VLBI scales should show a time dependence within a rotation period.
To further investigate this dependence, we divided the 2007 and 2010 VLBI observations into three time intervals, each corresponding to 5--6\,hr, which allowed us to obtain three "snapshot" images. The same procedure was applied to the 2013 data. However, only two snapshot images could be obtained for this epoch since shorter time intervals resulted in very sparse \textit{uv} coverage and, therefore, maps of degraded quality with unreliable structures.
The time covered by each snapshot is defined in Table~\ref{table:time_components}.
Fig.~\ref{fig:image_time} shows the snapshot maps 
corresponding to the same observational epoch (but different UT ranges) and the remarkable changes in the structure of AB~Dor~A. Epochs 2007 and 2010 start at similar rotational phases of the star \citep[with the epoch of zero rotation phase at JD 2444296.575;][]{1981A&A...104...33P}: 0.30 for 2007 and 0.39 for 2010, and, indeed, their snapshot images corresponding to the same 
time interval are similar, showing a double core-halo morphology, a feature 
that is also visible in the entire data set images.
However, the details of the structures change significantly on timescales of a few hours: the easternmost component seems to have
rotated (from snapshot 1 to snapshot 3) with respect to the brightest emission peak an angle of 40 $\pm$ 3$^{\circ}$ (47 $\pm$ 3$^{\circ}$) at epoch 2007 (2010).
We should emphasize that each snapshot image 
conserves its own astrometric information (referenced to the external quasar), that is, the snapshot images are properly registered. 
This allows us to measure the absolute motion of both components between snapshots 1 and 3, registering a total 0.6 mas (1.8 mas) movement for component 1 and 1.8 mas (2.0 mas) movement for component 2 in 2007 (2010).
We investigated whether these rotations of component 2 could be an artifact of the intrinsic rotation in the dirty beam.  As seen in Fig.~\ref{fig:image_time}, despite the similarity between images and dirty beams,  the position and rotation rate of the east side lobe and component 2 do not match. However, we should admit that some contamination may be present in our data, so our findings in terms of the time analysis should be taken with caution.

Finally, the appearance of the snapshots at epoch 2013 (see Figure~\ref{fig:image_time}) shows a very different behavior (although this epoch starts at a very similar phase to the 2010 data, i.e., 0.40) with significant changes between the two snapshot images:  the first snapshot, corresponding to the first half of the observation, shows a unique central component coincident with the brightest peak found in the maps constructed with the entire data set. Following the naming convention in Fig.~\ref{fig:allepochs}, this will be component 1a. 
However, in the second snapshot we recover the binary structure with the presence of component 2, being component 1a elongated due to component 1b. 
The absence of this double structure during the first half of the observations makes the behavior of AB~Dor~A in the 2013 epoch look different compared to that of 2007 or 2010. Significantly perhaps, the separation between components during the second half of 2013 also differs from the separation measured in 2007 and 2010 data (see Sect.~\ref{sect:imaging_and_modelfitting}).


     
    \begin{table}[b]
    \caption{Flux density of the components present in the snapshots (Fig.~\ref{fig:image_time}).} 
    \label{table:time_components} 
    \centering 
    \scalebox{0.95}{
    \begin{tabular}{ccccc} 
    \hline\hline 
    Epoch & Snapshot & $S_{\mathrm{1,1a}}$ & $S_{\mathrm{1b}}$ & $S_{\mathrm{2}}$ \\ 
    & & (mJy)& (mJy)&(mJy)\\
    \hline
    11 Nov. 2007 & & & &\\
    \quad 10:30-16:00 & 1 & 3.6 $\pm$ 0.3 &  &1.1 $\pm$ 0.2\\
    \quad 13:00-19:00 & 2 & 3.0 $\pm$ 0.5 &  &1.3 $\pm$ 0.3\\
    \quad 16:00-22:00 & 3 & 3.5 $\pm$ 0.6 & &1.5 $\pm$ 0.3 \\
    25 Oct. 2010 & & & &\\
    \quad 11:30-18:00 & 1 & 1.8 $\pm$ 0.3 &  &0.8 $\pm$ 0.3\\
    \quad 14:00-20:00 & 2 & 2.3 $\pm$ 0.4 & &1.1 $\pm$ 0.3 \\
    \quad 18:00-23:00 & 3 & 1.7 $\pm$ 0.4 &  &1.3 $\pm$ 0.4\\
    16 Aug. 2013 & & & &\\
    \quad 16:30-22:00& 1 & 5.4 $\pm$ 0.3 & &\\
    \quad 22:00-03:00& 2 & 1.5 $\pm$ 0.2 & 1.6 $\pm$ 0.3 & 2.3 $\pm$ 0.4\\
    \hline 
    \end{tabular}}
    \end{table}
    
    Following the same procedure described in Sect.~\ref{sect:data_red_and_analysis}, we fitted circular Gaussians to the interferometric visibilities for each time interval (Table~\ref{table:time_components}). 
    Due to the resemblance with the entire data set image and in order to make the comparison easier, we fixed the component sizes to those measured in Table~\ref{table:modelfit}. Both in 2007 and 2010, the brightest component flux remains constant (within errors) during the entire observation while the second component might be slightly increasing in flux during the second half. 
    Regarding epoch 2013, we notice that the flux density of the three components is, in some cases, larger in the snapshot maps (Table~\ref{table:time_components}) than that measured 
    in the entire data set image (Table~\ref{table:modelfit}), which is likely a consequence of the 
    time averaging over the complete observing time in the latter maps.
    
    
    
\subsection{Radio emission of AB Dor C}\label{sect:radio_abdorc}

    We found no emission at the expected position of AB~Dor~C, according to the well-known orbit of the system \citep{2017A&A...607A..10A} in any of our epochs. The non-detections place upper limits of 0.11 mJy, 0.04 mJy, 0.10 mJy, 0.04 mJy, and 0.07 mJy for the radio emission of this ultracool dwarf in 2007, 2010, 2013, 2017, and 2018, respectively.
    
    

\subsection{Orbit}\label{sect:orbit}


    We analyzed the orbital motion of the system in a similar fashion to \citet{2017A&A...607A..10A}, making use of previously reported VLBI, optical (\textit{HIPPARCOS}), infrared, and our new VLBI measurement (epoch 2018). We used a Bayesian approach with a model based on the definitions of \citet{2009ApJS..182..205W}. Our estimates of the astrometric and orbital parameters are in agreement with the values provided by \citet{2017A&A...607A..10A}. Indeed, we obtained dynamical masses of  0.91 $\pm$ 0.06 M$_{\odot}$ and 0.091 $\pm$ 0.005 M$_{\odot}$ for AB~Dor~A and AB~Dor~C, respectively.

\section{Discussion}\label{sect:discussion}
    
    

    There have been a few mechanisms proposed to explain the radio emission detected in active PMS stars like AB~Dor~A, as stated in the introduction. The usual explanation consists of synchrotron or gyrosynchrotron emission from large-scale magnetic structures or loops within a dipolar magnetosphere. \citet{1996AJ....111..918P} presented an alternative explanation for the case of HD 283447 (a binary system) modeled as a combination of smaller magnetospheres around each star and a common magnetic structure. These structures would be associated with the process of interacting magnetospheres that could form solar-like helmet streamers as reported by \citet{2008A&A...480..489M} in the young binary system V773~Tau~A, although this has not be confirmed by more recent observations \citep{2012ApJ...747...18T}.
    Recently, \citet{2018A&A...610A..81L} considered that the flares observed in some systems could be produced via the interaction between the magnetic field of the star and its close-by planets, although recent observations have not been able to find evidence of such interactions \citep{2016ApJ...830..107B}.
    Slingshots prominences might be another possible explanation for substructures around this star. However, although their presence in AB~Dor~A is well known \citep{1989MNRAS.236...57C}, these corotating clouds are mostly made of neutral gas that is not expected to produce emission at the observed frequencies. 
    
    As mentioned in Sect.~\ref{sect:imaging_and_modelfitting}, the radio emission detected in our 1.4 GHz image 
    could be interpreted as synchrotron or gyrosynchrotron emission that likely originated in relativistic electrons spiraling around the magnetic field lines in the outer layers of the corona. On the contrary, the 8.4 GHz images (Fig.~\ref{fig:allepochs} and Fig.~\ref{fig:image_time}) present a challenge to the models considered here since they must account for the following observed properties: 1) an internal structure consisting of compact features located at 10 R$_\mathrm{star}$ (in 2007 and 2010) and at 18 R$_\mathrm{star}$ (in 2013) away from the central component in the east direction; 2) variability of the components positions on a timescale of hours (see Fig. \ref{fig:image_time}); 3) low degree of circular polarization (<10\%); 4) brightness temperatures between 10$^7$ K and 10$^8$ K.
    
\begin{figure}
    \centering
    \includegraphics[width=1.0\linewidth]{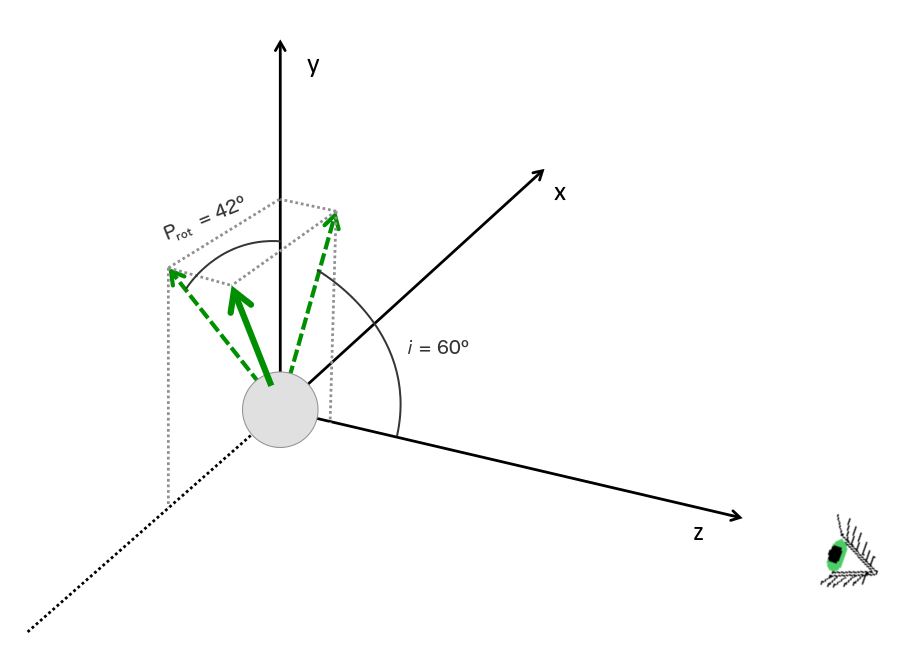}
    \caption{AB~Dor~A represented as a gray sphere with its rotation axis plotted in solid green, assuming the rotation axis is perpendicular to the orbital plane of the binary system AB~Dor~A/C. The dashed lines represent the projection of the rotation axis on the $xy$ plane (plane of the sky) and on the $zy$ plane. In this sketch, the rotation axis is inclined  $60^{\circ}$ with respect to our line of sight, here represented by the $z$ axis.
    }
    \label{fig:orientation}
    \end{figure}

    Any successful model must also be consistent with the dynamical configuration of the system, that is, with the well-known inclination of the rotation axis \citep[$\sim$60$^{\circ}$;][]{1994A&A...289..899K} with respect to our line of sight. 
We note that the position angle of the rotation axis in the plane of the sky, PA$_{\rm{rot}}$, is unknown. As a reference, we show in Fig. \ref{fig:orientation} the configuration resulting from a rotation axis perpendicular to the orbital plane of AB~Dor~A/C, as shown in Fig. 2 of \citet{2017A&A...607A..10A}, which corresponds to a position angle PA$_{\rm{rot}}$\,=42\,$^{\circ}$. However, there is no observational evidence supporting this particular value of PA$_{\rm{rot}}$; therefore, any orientation of the rotation axis in the plane of the sky can be assumed in the following discussion.
    
    On the other hand, the astrometric results of \citet{2017A&A...607A..10A} successfully predict the position of the brightest peak of our 8.4 GHz images (Fig.~\ref{fig:allepochs}). Given the scatter of 1.1 mas ($\sim$3 stellar radii) 
   of the astrometric measurements, it is reasonable to assume that the 
    peak of emission at each epoch corresponds to a radio emitting region that is near, but not necessarily coincident with, the stellar photosphere. Accordingly, the photosphere could be located several stellar radii away from the brightness peak. This is particularly relevant with regard to the discussion that follows.

\subsection{Polar cap hypothesis}\label{sect:polarcap}

    \begin{figure}
    \centering
    \includegraphics[width=\linewidth]{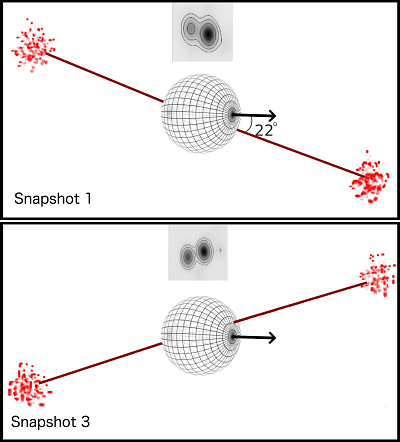}
    \caption{ 
    Polar-cap model applied to 2007 data: the rotation axis (black arrow) is inclined $60^{\circ}$ toward the observer but east-west oriented with PA$_{\rm{rot}}$$\sim$270$^{\circ}$; the magnetic axis (red line) is misaligned by $\sim$22$^{\circ}$ with respect to the rotation axis. 
    The rotation of the star would result in different orientations of the magnetic axis, effectively reproducing snapshot 1 (top panel) and snapshot 3 (bottom panel). The intermidiate stage would correspond to snapshot 2.}
    
    \label{fig:polar_model}
    \end{figure}

    The previous studies of Algol \citep{1998ApJ...507..371M} and UV~Ceti \citep{1998A&A...331..596B} revealed that a strong, large-scale, dipolar field could be consistent with the double-lobe structure observed in VLBI images of these objects. In these cases, each VLBI component would correspond to an emission 
    region located above each one of the polar caps of the star. In this model
    \citep[e.g.,][]{2002MNRAS.329..102K}, 
    electrons are accelerated at the equatorial region by the flaring activity of the star, then they move along the (dipole-like) magnetic field lines toward the magnetic poles.
    Our 8.4 GHz images (see Fig.~\ref{fig:allepochs}) posses a great morphological resemblance to the double-lobe structure detected in Algol and UV~Ceti. Hence,  a polar cap model may be consistent with our 8.4 GHz observations.
    
    Although the magnetic field of AB~Dor~A extrapolated from Zeeman-Doppler images is known to be much more complex than a simple dipole \citep[e.g.,][]{2010ApJ...721...80C}, the dipole-like contribution could also be significant in this star and polar emission similar to that of Algol and UV~Ceti could occur. Under this polar cap hypothesis, 
    the magnetic dipole axis must be oriented east-west to explain the preferred direction for the radio emission of Fig. \ref{fig:allepochs}. 
    Zeeman-Doppler images of the surface of AB~Dor~A showed that there is a misalignment between magnetic and rotation axes. Observations from 1995 to 2007 revealed that the latitude of the dipole component axis ranged from $-$35$^{\circ}$ to 57$^{\circ}$ indicating that this misalignment varies significantly over time \citep[see Table 1 in][]{2019MNRAS.tmp.2779J}. Therefore, it is plausible that the magnetic dipole axis was oriented east-west during our observations. 
 Under this assumption, the stellar disk in 2007 and 2010 (2013) would be located between components 1 (1a) and 2, which would be readily associated to radio emission above each of the polar caps at projected distance of $\sim$5 R$_{\mathrm{star}}$ ($\sim$9 R$_{\mathrm{star}}$) above the stellar surface, high enough to avoid, at least 
 partially, the emission becoming hidden by the stellar photosphere. 
   
    This polar-cap hypothesis can be further checked using the snapshot images of Fig.~\ref{fig:image_time}. To explain the temporal variation shown in these images, it is necessary to invoke a rotation axis oriented east-west (PA$_{\rm{rot}}\sim270^{\circ}$) with a misaligned magnetic axis, as sketched in Fig.~\ref{fig:polar_model}. According to this model, as the star rotates, emission coming from above the magnetic polar regions would create the different snapshots, which would explain  the relative motion between components 1 and 2 in epochs 2007 and 2010. 
    Snapshots at epoch 2013 would represent a special case. During the first  half of the 2013 observations, emission coming from only one of the  polar caps would be detected (component 1a) and during the second half of the observations, the emission from both polar caps would be detected.
    On the other hand,  if snapshots 1 and 3 of 2007 and 2010 were representative of the time of the maximum apparent separation between magnetic and rotation axes, then the estimated misalignment of the magnetic axis would be $\sim$22$^{\circ}$ for both 2007 and 2010. This misalignment is also compatible with the maps at epoch 2013. 

    The difference in flux between components might be either an intrinsic effect or a geometrical one, however, the fact that the brightest peak of emission permanently (at the three epochs) coincides with 
    the westernmost component may lend favor to a geometrical explanation for this effect. A possible scenario that would account for this difference in flux would be, as Fig.~\ref{fig:polar_model} shows, that the rotation axis of the star is inclined toward our line of sight so that polar emission coming from above this region would be directed toward us while emission from the opposite pole is pointing away from us, perhaps partially hidden, and thus detected as a weaker emission. This also may explain why only the westernmost component (polar cap pointing to the observer) is seen in the first snapshot of epoch 2013, although, certainly, the limited dynamic range of our snapshot images may not be enough to detect both sides of the emission (even more considering that component 1a was more than three times brighter in snapshot 1; see Table~\ref{table:time_components}). 
    

    
    Regarding the (absence of) polarization in our maps, we should mention that 
    both polarized and non-polarized polar-cap emission have been reported for other radio stars. It was found that Algol possesses two lobes of opposite circular polarization \citep{1998ApJ...507..371M}, whereas a low circular polarization was found in the pair of giant synchrotron lobes of UV~Ceti \citep{1998A&A...331..596B}. In principle, AB Dor A would be similar to the latter case, although further observations and a deeper understanding of the mechanisms involved in this type of emission may be necessary to clarify this point.
 
    Therefore, the polar cap hypothesis can explain our results considering 1) a rotation axis oriented east-west (PA$_{\rm{rot}}\sim270^{\circ}$ in the plane of the sky); and 2) a misalignment between the magnetic and rotation axis of $\sim$22$^{\circ}$. This is a fairly stringent, but certainly plausible scenario to explain 
    the complex internal structure of AB~Dor~A.

\subsection{Flaring loops hypothesis}\label{sect:loops}

    \begin{figure*}
    \centering
    \includegraphics[width=0.495\linewidth]{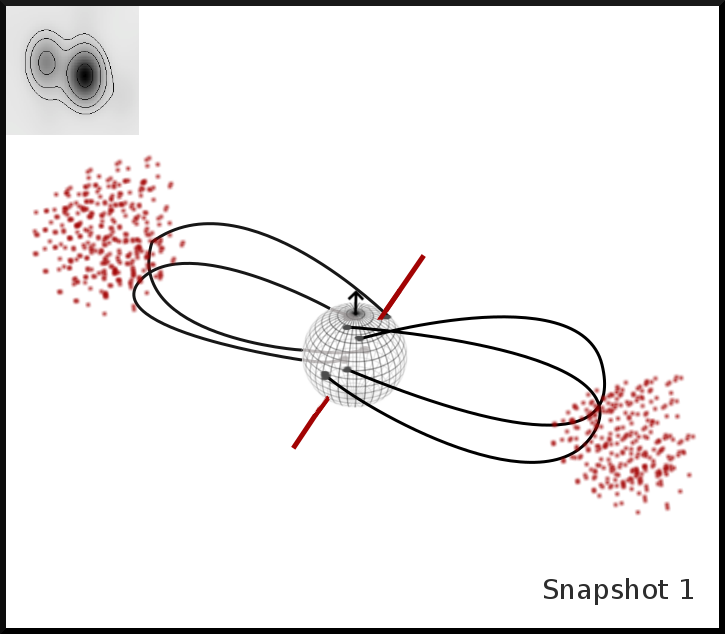}
    \includegraphics[width=0.495\linewidth]{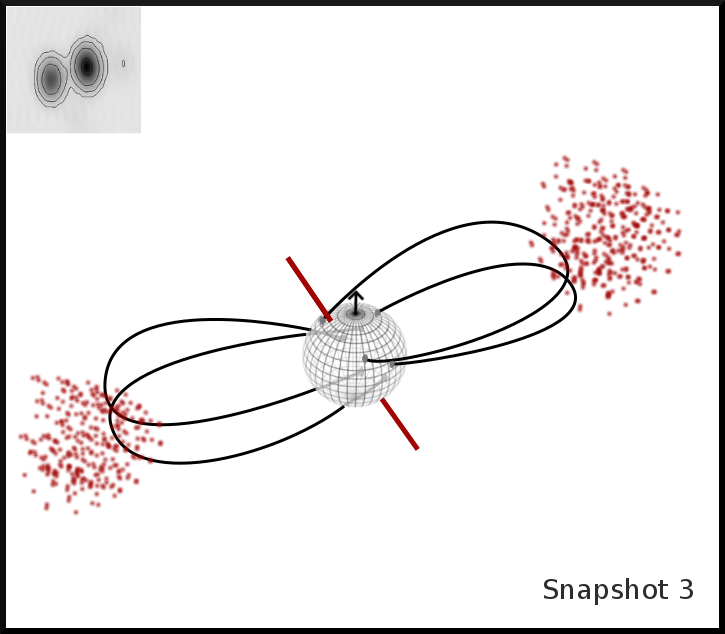}
    \caption{Flaring loop model applied to 2007 data. AB Dor A is represented as a sphere with  its rotation axis indicated by an arrow ($i\sim60^{\circ}$, PA$_{\mathrm{rot}}\sim0^{\circ}$) and a slightly misaligned magnetic axis ($\sim20^{\circ}$) which is plotted as a red line. The radio emission is originated by magnetic reconnection or interaction at top of the loop system surrounding the star.
    As discussed in the main text, the rotation of the star would create snapshot 1 (left panel) and snapshot 3 (right panel). 
    }
    \label{fig:loop_model}
    \end{figure*}
    

  A great number of flaring events are known to occur in AB Dor A \citep[see][and references therein]{2019A&A...628A..79S}. From these and previous studies, it has been well-established that the corona of this star presents evidence for both compact and extended structures. Typically, X-ray observations tend to detect small closed loops near the surface \citep{2000A&A...356..627M}, while H$\alpha$ absorption transients reveal cool condensations of neutral hydrogen trapped within the corona by the star magnetic field, extending several stellar radii away from the rotation axis of the star \citep{1989MNRAS.236...57C} and indicating the possible presence of giant loops in AB~Dor~A. On the other hand, \citet{2010ApJ...721...80C} carried out detailed numerical simulations and found that the corona of AB~Dor~A must be dominated by strong azimuthal tangling of the magnetic field due to its ultra-rapid rotation. This azimuthal wrapping of 
the magnetic field lines could lead to strongly curved, and eventually interacting, magnetic loops that may extend up to $\sim$10\,R$_{\mathrm{star}}$.  
Rather than episodic events (which are not excluded) and following the scenario proposed by \citet{1992ApJ...388L..27L}, these loops could constitute a quasi-permanent structure azimuthally distributed \citep[following the magnetic field wrapping as modeled by][]{2010ApJ...721...80C} where magnetic reconnection or interaction events may occur). These interactions would produce a permanent acceleration of the plasma electrons, which would justify the permanent radio emission.   The idea of an extended corona in AB Dor A is also justified by the above-mentioned presence of massive prominences at large distances from the stellar disk, associated, in turn, to the existence of large and energetic loops \citep[][and references therein]{2018MNRAS.475L..25V}. The morphological similarity between the images at 8.4 GHz in  Fig.~\ref{fig:allepochs} (east-west radio emission) certainly seems to support the presence of a long-lived, extended loop structure.

Under this hypothesis, component 1 and 2 in epochs 2007 and 2010 (Fig.~\ref{fig:allepochs}) would correspond, respectively,  to the east and west side of such a flaring structure at a height of $\sim$5R$_{\mathrm{star}}$ around AB~Dor~A (see Fig.~\ref{fig:loop_model} for easier visualization). We notice that a PA$_{\rm{rot}}$$\sim$$0^{\circ}$ and a slightly misaligned magnetic field 
($\sim$20$^{\circ}$) would successfully reproduce the snapshot maps in 2007 and 2010: 
considering the star rotation after 6 hours (time difference between snapshots 1 and 3)  the magnetic axis would have rotated $\sim$180$^{\circ}$ around the rotation axis which, translated to our images, would correspond to the different orientation of the emission seen between snapshot 1 and 3 (see Fig. \ref{fig:loop_model}). 
This model is also compatible with the snapshot images in 2013, assuming that the easternmost (in principle, weaker) side of the flaring structure is not seen or detected during the first part of the observations. 
We notice that the registration of the snapshot images seems to support this hypothesis as both component 1 and 2 (in 2007 and 2010) move in opposite directions (see Fig.~\ref{fig:image_time}) as it would correspond to the motion of the azimuthally-distributed emission region due to the rotation of the misaligned magnetic field axis.
On the other hand, the question of why component 1 is permanently brighter than component 2 does not 
have a unique answer in this model, although it could be explained by a combination of 1) inhomogeneities in the radio emission region; 2) orientation effects of a clockwise rotation, given the high directivity of the radio emission (as corresponds to synchrotron emission); and 3) occultation or absorption due to the stellar disk or slingshot prominences.

\subsection{Helmet streamers hypothesis}\label{helmet_streamers}

    \begin{figure}
    \centering
    \includegraphics[width=\linewidth]{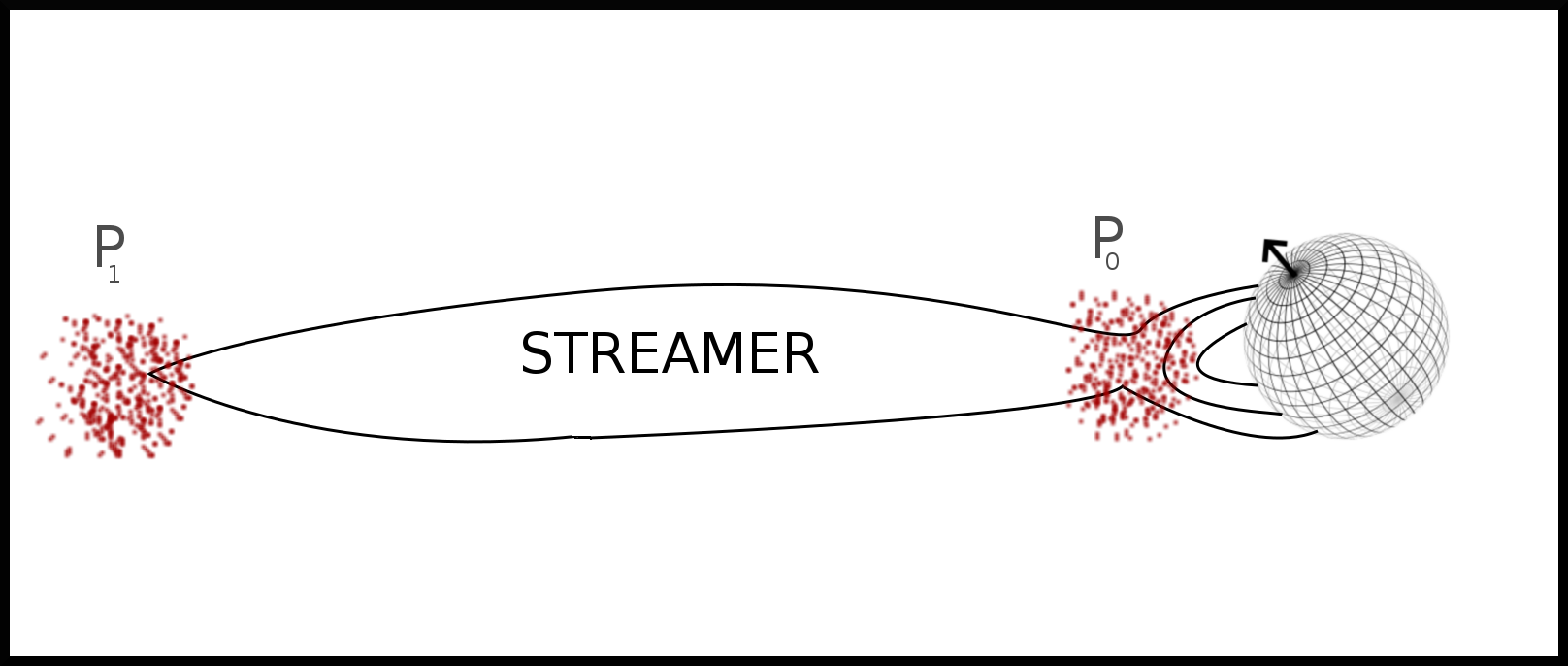}
    \caption{Sketch of the emission and magnetic structure of a helmet streamer in AB~Dor~A. P$_0$ and P$_1$ denote the lower and upper mirror points, respectively, where the radio emission is generated.  }
    \label{fig:helmet_streamer}
    \end{figure}

    As introduced in the previous section, the slingshot prominences detected by \citet{1989MNRAS.236...57C} have been shown to be corotating condensations of mostly neutral gas whose physics has been abundantly studied \citep{2000MNRAS.316..647F,2005MNRAS.361.1173J,2019MNRAS.482.2853J,2019MNRAS.483.1513W,2019MNRAS.485.1448V,2019MNRAS.tmp.2779J}. Similarly to the solar case, helmet streamers and flares are expected to occur at the top of such prominences (although not exclusively) triggered by magnetic reconnection of the coronal loops. 
    Evidence of stellar helmet streamers has been reported for the binary system V773~Tauri~A via VLBI observations \citep{2008A&A...480..489M}, which revealed a complex magnetic structure around this object, although more recent observations did not confirm such structures \citep{2012ApJ...747...18T}.
    The well-known presence of frequent slingshot prominences at large height in AB Dor A led us to consider that its internal mas-scale structure in radio could be the observational evidence of the formation of helmet streamers on top of magnetic loops. Supporting this hypothesis, and only one month after our 2007 observations, \citet{2019MNRAS.tmp.2779J} reported two unusually large slingshot prominences in AB Dor A, which may have been the consequence of a very energetic event. Although the lifespans of these phenomena are 2--3 days,  the formation and ejection of prominences are thought to be part of a continuous process in AB~Dor~A \citep[up to 18 events per day;][]{2018MNRAS.475L..25V} and the presence of associated helmet streamers would also be. This recurrent scenario would justify the permanent extended structure seen in Fig.~\ref{fig:allepochs}. 
    
Within a helmet streamer, confined particles travel along the streamer and are reflected back between both lower and upper mirror points \citep[see Fig.~\ref{fig:helmet_streamer};][]{1976MNRAS.176...15M}, both of which would constitute the radio emitter regions. Based on this hypothesis, component 2 in our images of epochs 2007 and 2010 would be interpreted as the upper mirror point, meanwhile the brighter component 1 would correspond to the lower mirror point blended together with the coronal emission produced near the stellar disk. In the particular case of epoch 2013, the presence of the streamer would be detected only during the second half of the observation, which would create component 2.
As can be seen in  Fig.~\ref{fig:allepochs}, the streamer would not corotate with the star, otherwise we would have detected its emission coming from the west-side of the star during the 12 hours of observation. This is reinforced by our snapshots images (Fig.~\ref{fig:image_time}), which show that over the duration of the entire observation, component 2 had an apparent rotation of only 40$^{\circ}$ (47$^{\circ}$) in 2007 (2010) between snapshot 1 and 3, separated by 6 hours. 

Is a non-corotating helmet streamer an acceptable scenario? According to our images, the streamer would extend up to 9--10~R$_{\mathrm{star}}$ (18 R$_{\mathrm{star}}$) in 2007 and 2010 (2013). These values are considerably lower than the Alfv\'en radius (distance below which the magnetic field dominates the gas pressure and forces the material to co-rotate with the star) for AB Dor A, which is estimated to be 24\,R$_{\mathrm{star}}$ \citep{2018MNRAS.475L..25V}. These values do not favor a  
non-corotating streamer in AB~Dor~A.  
However, there is no clear evidence that helmet streamers must co-rotate with the star; actually, the observed positions of the tentative helmet streamers in V773~Tau~A do not correspond to those predicted by the stellar rotation \citep{2008A&A...480..489M}. After all, once the helmet streamer has been produced, it is no longer attached to the star surface and may be subject to the curve following the strongly wrapped magnetic field of AB~Dor~A. Even more, in correspondence with the continuous ejection of slingshot prominences, the upper mirror points may also be expelled due to the fast rotation. 
However, the preference for eastward helmet streamers at the three epochs would have an unclear origin.

\subsection{Close companion hypothesis}\label{sect:companion}

    Although it may be tempting to interpret the 8.4 GHz images (Fig.~\ref{fig:allepochs}) as a binary system (identifying AB~Dor~A as component 1 and the companion as component 2), the temporal analysis of Fig.~\ref{fig:image_time} makes this scenario highly unlikely. Assuming that the axis of the orbital plane is parallel to the rotational axis of AB~Dor~A (left panel in Fig.~\ref{fig:orientation}),
    at only 3 mas of separation, the fast orbital motion of component 2 (both in 2007 and 2010) would imply 
    unacceptably large values of the radial velocity semi-amplitude of the stellar 
    reflex motion, much greater than the precision of previously-reported radial velocity 
    measurements of AB Dor A \citep[32.4 $\pm$ 2.2 km·s$^{-1}$;][]{2006AstL...32..759G}.
    

\subsection{AB Dor C}\label{sect:nodetectionC}
    
    Our non-detections at any of the observed frequencies put a strong upper bound on the flux density of the ultracool dwarf AB~Dor~C (see Sect.~\ref{sect:radio_abdorc}).
    The study of samples of ultracool dwarfs \citep{2012ApJ...746...23M,2013ApJ...773...18R} has shown that these kind of objects are not typically expected to exhibit radio emission. Nonetheless, some authors have detected radio emission from late M, L and T objects \citep{2013PASP..125..313M,2015Natur.523..568H,2018A&A...610A..23G} which has been associated with an electron cyclotron maser emission mechanism. They might have benefited from the strong correlation between auroral radio emission and the presence of a H$_{\alpha}$ line \citep{2016ApJ...818...24K}. In spite of the fact that there is no current evidence for H$_{\alpha}$ line emission in AB~Dor~C, there are some facts that might indicate radio emission in this ultracool dwarf: 
    \begin{tasks}
        \task young and late M-dwarfs, such as AB~Dor~C, tend to present H$_{\alpha}$ emission \citep{2008PASP..120.1161W} which may be indicative of auroral radio emission.
        \task The model of \citet{2010A&A...522A..13R} and the scaling law reported in \citet{2009Natur.457..167C} (magnetic field $\propto$ energy flux, valid for fully convective, rapid rotating objects) predict magnetic fields >$10^{3}$ G for an object with mass as 
        low as 72 $M_J$.
        \task AB~Dor~A is known to possess a strong magnetic field and H$_{\alpha}$ emission \citep[][and references therein]{2013A&A...560A..69L}. Since the system could have been formed from the collapse and fragmentation of the same rotating cloud, AB~Dor~C may have retained some of the characteristics of AB~Dor~A, that is, high rotational velocity and a strong magnetic field. These features could then be responsible for radio emission in this ultracool dwarf, although it would be variable in time or fainter than expected, which would explain our non-detection.
    \end{tasks}
        More sensitive observations would be necessary to address the possible radio emission of this object. The radio detection of AB~Dor~C would be of great importance in probing its emission mechanism and could provide some insights into its structure.
        
\section{Conclusions}\label{sect:conclusion}

    Our multi-epoch, multi-frequency observations of the binary system AB~Dor~A/C revealed an intriguing scenario in the main star of the system. At 1.4 GHz, AB~Dor~A was detected as a compact source with no flux variability during the 12-hour  period of our observation and with a circular polarization lower than 10\%. With a minimum brightness temperature of $3.5\times10^{7}$~K, the origin of this emission seems to be non-thermal gyrosynchrotron or synchrotron emission coming from accelerated electrons located in the outer layers of the corona. A tentative detection (with S/N~$\sim$~5) was also made at 22.3 GHz with emission located $\sim$18 R$_\mathrm{star}$ away from the expected position, and whose origin remains unclear. A re-analysis of the 8.4 GHz data resulted 
    in the detection of an extended and variable structure, morphologically similar at all epochs. We considered four different scenarios in order to explain the 8.4 GHz observations:
    
    

        
        
    
    \begin{enumerate}
        \item Polar cap model with emission coming from above the polar regions of the star. Under this hypothesis, the rotation axis would be oriented east-west, and the magnetic axis would be slightly misaligned with respect to the rotation axis. This model is capable of successfully explaining the 2007 and 2010 preferred east-west orientation of the emission, the relative motion between components, and the fact that the western component seems to be always the brightest. One of the polar cap regions is not seen (or not detected) in the first snapshot image of epoch 2013. Although the geometric requirements are stringent, this model explains the observational features seen in our 8.4 GHz images. 
       
        \item Flaring loop model where the emission would originate due to magnetic reconnection of a coronal loop structure distributed azimuthally following the magnetic field lines, which would, in turn, be wrapped by the rapid rotation of AB~Dor~A. On the basis of a nearly north-south rotation axis, a slightly misaligned magnetic axis would rotate, reproducing the different 
        snapshot images with the emission coming from magnetic reconnection events at heights 5--9 R$_{\mathrm{star}}$.  Again, in the case of 2013, during the first half of the observation, only one side of the loop structure was detected. It is not clear how 
        this model may justify why the westernmost component is the brightest one, although a combination of orientation and absorption effects could produce such a result.

        
        \item Helmet streamer model where the radio emission is originated at the upper and lower mirror points, the latter likely merged with coronal emission of AB~Dor~A itself. Helmet streamers are, in principle, associated to large and continuously produced slingshot 
        prominences; our maps suggest that the streamers may be curved due to fast rotation and may not necessarily corotate with the star. 
        Helmet streamers could successfully explain most of the observed characteristics of our data; however, the preferred eastwards orientation of the radio emission needs to be properly explained under this model.
        
        \item Scenario including a close companion to AB~Dor~A, which would readily justify the binary radio morphology. This model is, however, highly unlikely due to the small separation between both components which would 
        produce unacceptable large radial velocity values, a result that is incompatible with previously published observational measurements of AB~Dor~A.
    \end{enumerate}
    
    Whichever model is correct for AB~Dor~A, our results confirm the extraordinary coronal magnetic activity of this star, capable of producing compact radio structures at very large heights which have so far only been seen in interacting binary systems.
    Lastly, no emission was found at the expected position of the ultracool dwarf AB~Dor~C, placing strong upper limits on this binary brown dwarf. New data will provide an excellent opportunity for furthering investigations of this remarkable system.
    
\begin{acknowledgements}
J.B.C., R.A., J.C.G., and J.M.M. were
partially supported by the Spanish MINECO projects AYA2012-38491-C02-01, AYA2015-63939-C2-2-P, PGC2018-098915-B-C22 and by the Generalitat Valenciana projects PROMETEO/2009/104 and PROMETEOII/2014/057
\end{acknowledgements}

\bibliographystyle{aa_url} 
\bibliography{references}

\end{document}